\documentclass[%
reprint,
superscriptaddress,
nofootinbib,
 amsmath,amssymb,
aps,
prx,
floatfix,
]{revtex4-2}

\usepackage[titles]{tocloft}
\usepackage[T1]{fontenc}
\usepackage{graphicx}
\usepackage{dcolumn}
\usepackage{bm}
\usepackage[hypertexnames=false]{hyperref}
\usepackage{siunitx}
\usepackage{braket}
\usepackage{rotating}
\usepackage{algorithm} 
\usepackage{algpseudocode} 
\usepackage{xfrac}
\usepackage{etoolbox} 
\usepackage{array}
\usepackage{enumitem}
\usepackage{eurosym}

\cftsetindents{section}{0em}{2em}
\cftsetindents{subsection}{2em}{2em}
\cftsetindents{subsubsection}{4em}{2em}

\usepackage[dvipsnames]{xcolor}
\usepackage[sectionbib]{bibunits}
\defaultbibliographystyle{apsrev4-2} 

\usepackage{hyperref}
\hypersetup{
    colorlinks=true,
    linkcolor=blue,
    urlcolor=magenta,      
    citecolor=Cyan,
    }
\renewcommand{\selectlanguage}[1]{}
\patchcmd{\subsubsection}{\itshape}{\bfseries}{}{}

\begin{document}
\begingroup
\setlength{\skip\footins}{0.5cm}
\title{Quantum Internet Use Case Analysis for the Automotive Industry}

\author{K. L. van der Enden}
\thanks{These authors contributed equally to this work \newline QuTech correspondence to: \href{mailto:publications@interseqt.nl}{publications@interseqt.nl}}
\affiliation{QuTech \& Kavli Institute of Nanoscience, Delft University of Technology, 2628 CJ Delft, The Netherlands}

\author{R. Kirschner}
\thanks{These authors contributed equally to this work \newline QuTech correspondence to: \href{mailto:publications@interseqt.nl}{publications@interseqt.nl}}
\affiliation{Formerly at Porsche Digital GmbH, Stralauer Allee 12, 10245 Berlin, Germany}
\affiliation{Currently at Audi AG, Auto-Union-Str. 1, 85057 Ingolstadt, Germany}

\author{M. Krumt\"unger} 
\affiliation{Formerly at Porsche Digital GmbH, Stralauer Allee 12, 10245 Berlin, Germany}

\author{A. Wilms}
\email[Porsche Digital correspondence to: ]{alissa.wilms@porsche.digital}
\affiliation{Porsche Digital GmbH, Stralauer Allee 12, 10245 Berlin, Germany}
\affiliation{Dahlem Center for Complex Quantum Systems, Freie Universität Berlin, 14195 Berlin, Germany}

\date{\today}
\begin{abstract}
A future quantum internet brings promising applications related to security, privacy and enabling distributed quantum computing. Integration of these concepts into the future trends of the automotive sector is of considerable interest, as it enables both the development of practical quantum internet use cases and the adoption of innovative technologies in the automotive sector. In this work we analyze cross-platform megatrends in both the quantum internet and the automotive industry, identifying mutually beneficial regions of interest. In the short-term ($<10$ years) hardware miniaturization and automation of quantum internet technology provides a synergy interface between the two domains. For the long-term ($\geq10$ years) we develop a comprehensive list of use cases for the quantum internet within the automotive sector. We find considerable relevancy of augmenting autonomous driving, vehicle ad hoc networks and sensor fusion with blind quantum computing, anonymous transmission and quantum cryptographic tools. These results can be used to target future research, engineering and venture developments for both domains. Furthermore, our approach can be applied to other industries, enabling a structured methodology for identifying and developing feasible use cases for the quantum internet in diverse domains.
\end{abstract}
\maketitle
\begin{figure*}[t]
    \centering
    \includegraphics[width = \textwidth]{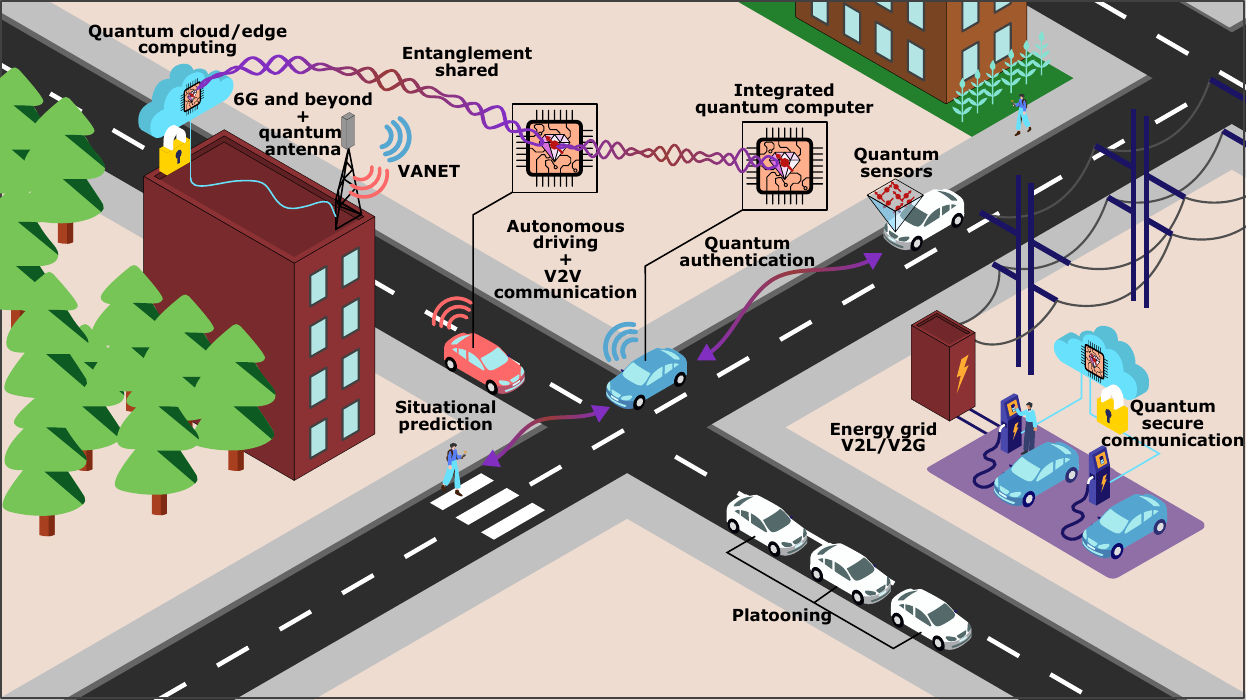}
    \caption{\textbf{Future quantum enabled automotive}. A sketch of an envisioned future of the automotive integrated with quantum technologies. Cars have integrated quantum computers on board, entangled with other cars and quantum (edge) servers. Entanglement is either generated live while driving via quantum antennas that operate in parallel with 6G networks and beyond, or pre-loaded in quantum memories while charging. Charging infrastructure has intelligent load balancing for vehicle-to-load/grid (V2L/V2G), where cars quantum securely communicate, use quantum digital signatures in their handshake, allowing payment to be processed.  Vehicles drive autonomously individually or together in platooning order. They set up vehicle-to-vehicle ad-hoc networks (V2V, VANET) to make driving decisions, aid in situational prediction and verify identities via quantum authentication. Cars can be equipped with quantum sensors to aid positional tracking for self-driving capabilities.}
    \label{fig:qi_automotive_future_cartoon}
\end{figure*}

\tableofcontents
\section{Introduction}
The automotive sector has been progressing towards integrating computational, communication and sensor technology in vehicles to enhance the user experience and strengthen safety measures, among many other improvements~\cite{NHTSA_safety, infotainment_history}. Current trends in the automotive sector require the integration of cutting-edge technology, such as AI, 5/6G networks and batteries to enable automated driving, over-the-air (OTA) software updates and electrification~\cite{Abdelkader2021}. The introduction of these and further new technologies may make vehicles a higher risk to be vulnerable to cyber attack vectors, which is an active area of concern~\cite{Checkoway2011, Olufowobi2019, Khan2020, Takahashi2020, pekaric2021taxonomyattack,  Renganathan2022}. In parallel, the last decades have shown considerable advancement in the quantum technology research fields of quantum computing, quantum internet and quantum sensing. These technologies promise to solve particular complex computational problems~\cite{Shor1994, Grover1996, Lloyd1996, Preskill2018}, provide quantum-secure communication and ability for networked quantum computing~\cite{Kimble2008, Wehner2018}, and enhanced sensing~\cite{Campbell2017, Degen2017, Pirandola2018}, respectively.  Quantum technologies have the potential to provide the innovation required to satisfy the increased needs in technological capabilities and security requirements for the envisioned future trends in the automotive sector. In addition, due to their novelty, applications of quantum technologies could bring forth use cases within the automotive sector that previously had not been considered. 
\vspace{\baselineskip}

Current quantum and automotive technological development and use cases mostly manifested in the field of quantum computing, focused on algorithms for traffic routing and factory process optimization, supporting machine learning approaches for autonomous driving and improved batteries for electric vehicles~\cite{Yarkoni2020, burkacky2020will, Bayerstadler2021, yarkoni2021multi}. 
Conversely, little attention has been given to exposing the automotive domain to the possible applications that the quantum internet provides.
However, with a future quantum internet having promising applications in secure and privacy enhancing communication and enabling distributed quantum computing, the benefit of employing these concepts in the automotive sector can be substantial. Furthermore, the automotive domain works on development cycles that make for a natural match with quantum research topics, of which many are on a decade timeline towards being a commercial product. Reciprocally, the quantum internet domain can benefit from an additional source of practical applications for the technologies they develop. 

\vspace{\baselineskip}
Given the current state of the art, it remains uncertain which, if any, of these potential applications could be realized in practice. This analysis explores their potential of implementation under the assumption of an optimal scenario. We utilize a method where megatrends - trends that have an effect on the entirety of their respective industry - of both fields are identified and qualitatively compared for both short ($<10$ years) and long ($\geq10$ years) time scales. First, we will cover the need-to-know of the quantum internet and its relevant applications in Section~\ref{section:intro_qi}. We then introduce the automotive field in Section~\ref{section:intro_automotive} and show why these two sectors are well matched for developing mutual use cases. Next, we will describe the megatrends of both respective sectors in Section~\ref{section:megatrends}. We investigate the long and short term trends of the quantum internet in Section~\ref{section:qi_applications} and~\ref{section:qi_research_trends}, respectively. We repeat this for the automotive for the long and short term trends in Section~\ref{section:megatrends_autonomous_future_trend} and~\ref{sec:megatrends_automotive_prePDP}, respectively. Then, we perform a qualitative comparison of these trends and underlying sub-trends in terms of their mutual interfacing and developmental relevance in Section~\ref{section:qi_prePDP}. On this analysis we perform a synergy evaluation in Section~\ref{section:synergy_evaluation}, allowing us to extract the trends that have the largest potential to build business cases around. Finally, we show newly developed potential use cases for the future of the automotive sector and make a qualitative assessment of the potential relevance of quantum internet applications within these use cases in Section~\ref{section:automotive_future_trends_qi_applications}. 
\endgroup

\subsection{Quantum Internet}
\label{section:intro_qi}
Analogous to how the current (classical) internet enabled a revolution for the classical computer, interconnecting quantum computers promises a similar revolution to quantum computing. Such a network of interconnected quantum computers is a quantum internet. Unlike a solely classical network where information is transmitted using classical bits, a quantum internet will transmit quantum bits (qubits) between quantum computers. Due to their quantum nature, these qubits cannot be sent over the classical internet, thus a different method of transmission of quantum information is required. A key property that enables sharing of quantum information is quantum entanglement. Entanglement between qubits can be generated with various methods and for telecommunication compatibility, generally involve using so-called flying qubits, single particles of light (photons). Once entanglement is generated, it can be consumed to teleport quantum information~\cite{bouwmeester1997experimental, rodney_ch4, rodney_ch4}, the quantum equivalent of sending classical information. In contrast to the classical, quantum information cannot be copied (no-cloning theorem~\cite{Wootters1982}), it disappears at the sender and appears at the receiver. In part, the property of the no-cloning theorem allows detection of when outside intervention on quantum communication is attempted, making it an opportune technology for verification of security of communication channels~\cite{Bennett2014}. The downside of this property is that amplification of signals is impossible, and repetition of quantum information is not straightforward - one must employ the use of specialized quantum repeaters that use similar underlying quantum properties to mediate long-distance ($>\SI{100}{\kilo\meter}$) communication~\cite{rodney_ch10}.

A particular application of entanglement is to securely generate classical encryption keys between different parties using entangled photons. This is known as Quantum Key Distribution (QKD)~\cite{Ekert1991, Bennett2014, rodney_qkd}, a technology that is available commercially and implemented worldwide~\cite{stanley2022recent, idquantique, toshiba, magiq, quintessencelabs}. The usage of QKD is to generate these classical encryption keys that are secure against classical and quantum attacks, in contrast to current quantum unsafe cryptographic protocols (RSA, AES)~\cite{Shor1994, Grover1996}. While currently accepted post-quantum cryptographic (PQC) algorithms are classical and quantum secure~\cite{pqc}, they are not proven to be so indefinitely. QKD offers the alternative to be both currently and in the future provably classical and quantum secure. Still, the functionality of QKD ends with key generation, whereas the previously defined quantum internet has the ability to go beyond that - it has the ability to let quantum computers communicate in a network. The rest of the paper will mostly focus on the latter capabilities.

Complex and widely rolled out quantum networks have yet to be built, however there is widespread confidence that such networks will come to exist in the future, which is strengthened by considerable public funding in pursuit of that goal~\cite{QIA_phase1}. Currently, the cutting-edge of quantum networks allows for two and multi-node networks within research lab environments~\cite{Humphreys2018, Pompili_multinode_2021}, and successful advances are made in building different kinds of quantum networks at metropolitan scales using commercially available fiber networks~\cite{Liu2023, Rakonjac2023, Stolk2024}.  The quantum internet has use cases defined at different levels of functionality in the developmental process towards these complex networks~\cite{Wehner2018}. We continue discussing the relevant functionality levels and quantum internet applications for this paper in Section~\ref{sec:megatrends_QI}.

\subsection{Automotive}
\label{section:intro_automotive}
The global vehicle manufacturing industry is over a century old and reached \$2.6 trillion in revenue in 2023~\cite{car_revenue}. It employs in the EU almost 13 million people indirectly, of which 2.4 million directly~\cite{acea_jobs_2023}, making it a cornerstone of the economy covering around 7\% of all jobs in the EU. Technological developments are constantly integrated into new automobiles to introduce new or improved features for customers, or to adhere to changed regulations. As such, the spending of the automotive industry on R\&D is the highest of all industrial sectors in the European Union with \euro59 billion in 2021~\cite{rd_automotive_2021}, and a significant contributor to technology patenting with 587k patents filed between 2010 and 2019 by the top 20 world patent holders in the automotive industry~\cite{hafner2020european}.

The automobile integration of technology accelerated in the information age driven by regulations, user demands and the pursuit of technological advancements, resulting in cars have increasingly become more of a computer on wheels~\cite{Broy2006, Lim2011, Greengard2015, Macher2015, navet2017automotive, burkacky2019automotive}. User interfaces are on (touch)screens, control of automotive functionality is relinquished to digital processing and car access can be keyless, to name a few. This trend continued with the introduction of sensory-awareness and driving assistance for enhanced safety and driver comfort, aiming towards full autonomous driving in any situation~\cite{sae_autonomous_driving}. There are many challenges to be solved in achieving that goal, not only involving the self-contained challenge of an autonomous driving vehicle interacting in traffic, but also in the security of data-transmission of sensitive information. In addition, these systems will need to be robust against rogue actors trying to gain access to driving systems individually or to vehicle fleets~\cite{Olufowobi2019, Malik2020}.

Furthermore, new vehicles are now connected to the internet. This enabled improved customer functionality for navigation and entertainment and enabled the possibility for vehicle-to-everything (V2X) communication, which includes over-the-air (OTA) diagnostics and updating of soft- and firmware, and smart-grid integration~\cite{Lu2014, vanet_Farsimadan2021, Abdelkader2021}. However, cybersecurity is becoming a key focus point for vehicle system development that permeates through the entire software stack of vehicle and V2X management software~\cite{Khan2020, Malik2020}.

To uphold extensive and thorough safety standards, vehicles must be compliant with regulations such as International Organization for Standardization (ISO) 26262\footnote{Road vehicles - Functional safety~\cite{iso26262}}. Furthermore, technology in cars aims to be robust and reliable, requiring only intermittent maintenance for eroding or moving parts.
These design requirements have impact on integrating new technologies in the Product Development Process (PDP) of automobiles, allowing only the introduction of new features after rigorous tests and verification of long-term operability. The phase before starting the PDP is the pre-Production Development Process (pre-PDP). In this initial stage, essential requirements are outlined to determine whether a new technology can be successfully integrated into the vehicle~\cite{PDP2015}. These pre-PEP requirements are shared between the vehicle manufacturer and their suppliers to form a mutual framework for product integration.

To formulate pre-PDP requirements for this paper, we extract the requirements set in the selection matrix for sensor component requirements from~\cite{winner2011handbuch}, generalizing it as a method for technology transfer of a new technology to the vehicle ecosystem, which makes it applicable to quantum technologies.  We present a non-exhaustive list of several of the most important requirements that are relevant to this analysis and discuss them further in Section~\ref{sec:megatrends_automotive_prePDP}.

In general, PDP cycles from concept to start of production can take up to 5 years~\cite{filipovic2014npd}. The automotive sector thus has to predict and steer future technological trends for their purpose beyond this 5 year timeline to evaluate their benefits, limitations and determine necessary steps for automotive-specific integration. It is this timeline that makes for a natural match with quantum research topics, of which many are on a decade timeline towards being a commercial product. This makes it an interesting candidate for matching pre-PDP requirements to current ($<10$ years) quantum internet development tracks. In addition, the timeline of predicting long term ($\geq 10$ years) use cases in the automotive sector aligns with the long term vision of quantum internet applications gaining usefulness as technological functionality increases. This allows for analyzing the intersection of future use cases in the automotive with applications of the quantum internet, which is discussed in  Section~\ref{section:automotive_future_trends_qi_applications}.

\section{Megatrends}
To conduct a thorough use case analysis, we need to identify key trends in both the quantum internet and
automotive industries for the long and short term. Trends that have an effect on the entirety of their respective industry are called megatrends. We discuss long and short term trends of the quantum internet in Section~\ref{section:qi_applications} and~\ref{section:qi_research_trends}, respectively. We repeat this for the automotive for the long and short term trends in Section~\ref{section:megatrends_autonomous_future_trend} and~\ref{sec:megatrends_automotive_prePDP}, respectively. We present the results of the interfacing analysis of the megatrends in Section~\ref{section:interfacing_analysis}.

\label{section:megatrends}
\begin{figure*}
    \includegraphics[width = 0.75\textwidth]{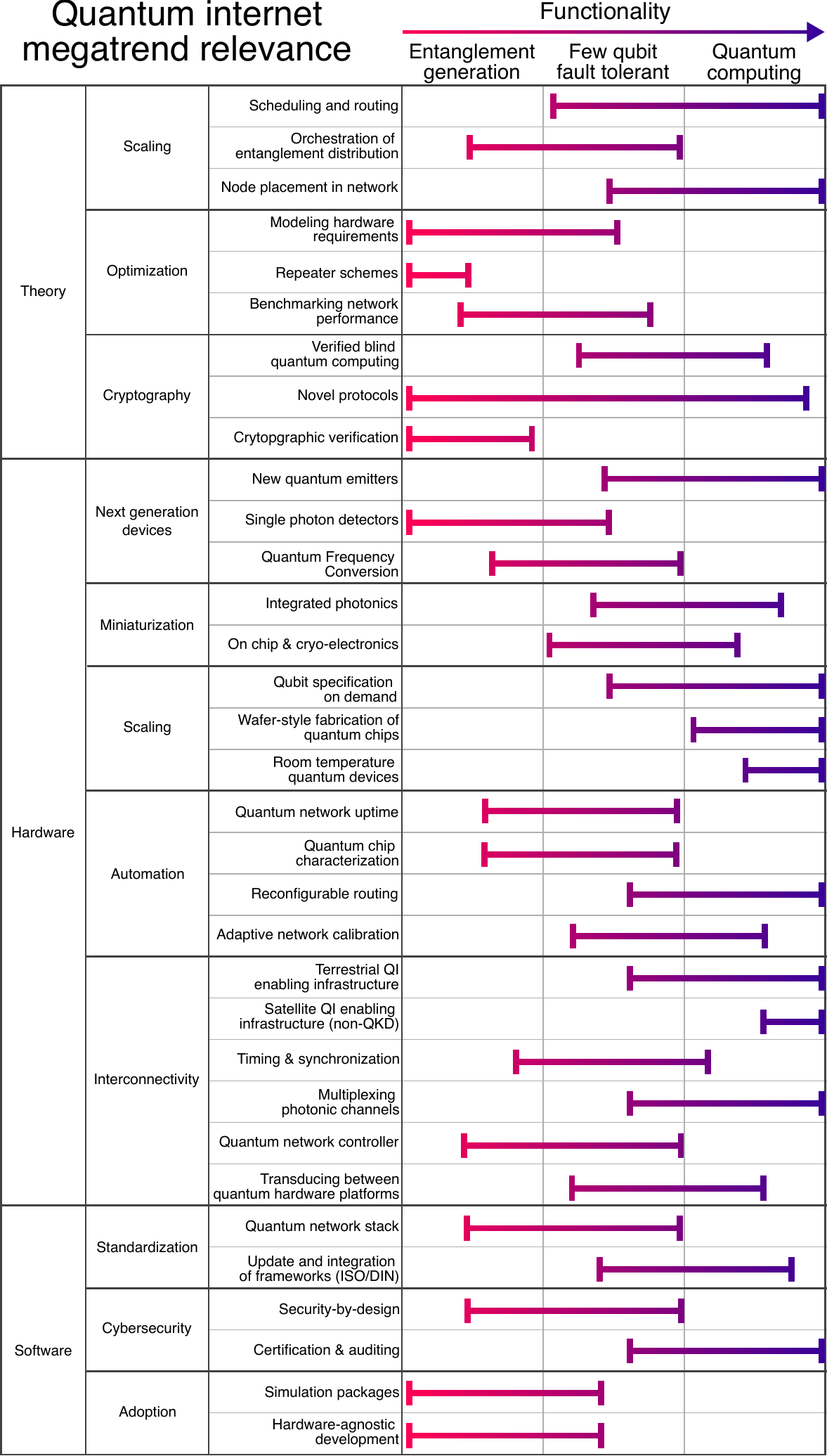}
    \caption{\textbf{Quantum internet megatrends.} List of quantum internet trends divided by category. We evaluate when this trend will be relevant at what functionality level of the quantum internet.  }
    \label{fig:qi_megatrends}
\end{figure*}

\subsection{Megatrends Quantum Internet}
\label{sec:megatrends_QI}
The quantum internet is at an early developmental stage, enabling us to identify two megatrend directions: Trends for future applications ($\geq 10$ years)~\cite{Wehner2018} and known protocols~\cite{singh2023unified} on one hand, and trends in current research tracks ($<10$ years) on the other. The latter are trending research tracks that are pursued to reach the future applications.

\subsubsection{Quantum Internet Future Trends}
\label{section:qi_applications}
 We can infer future application megatrends through recent research publications and guidance brought by institutional and industrial consortia~\cite{QIA_phase1, Awschalom2021}. From this, we distinguish three categories that we discuss below:

 \begin{enumerate}[noitemsep, label=\alph*.]
     \item \textit{Interconnected quantum computing}
     \item \textit{Quantum cryptography}
     \item \textit{Quantum sensors}
 \end{enumerate}
The trends identified are highlighted in bold in the text.

\vspace{\baselineskip}
\begingroup
        \centering
        \paragraph{Interconnected quantum computing}
        \mbox{}\\
\vspace{\baselineskip}
\endgroup
\label{sec:interconnected_quantum_computing}
 Interconnecting quantum computers is the leading trend in the field. With the knowledge that interfacing classical computers brought a second revolution to the information age, the quantum internet has the possibility to provide a similar revolution to quantum computing. Hence, \textbf{distributed quantum computing} is a sought after application that could accelerate reaching useful quantum computing. Furthermore, networked quantum computing capabilities also allow for \textbf{blind quantum computing} (BQC)~\cite{Broadbent2009, Fitzsimons2017}, where clients can run calculations on a server that will remain blind to what the client executes. This is expected to become an important cornerstone for future privacy and geographical-safe computing. Lastly, \textbf{anonymous transmission} of quantum information can provide built-in privacy for quantum networks~\cite{Christandl2005, Bouda2007, Unnikrishan2019}. 
 
\vspace{\baselineskip}
\begingroup
        \centering
        \paragraph{Quantum cryptography}
        \mbox{}\\
\vspace{\baselineskip}
\endgroup
\label{sec:quantum_cryptography}

A separate category of quantum communication focuses specifically on quantum-enabled cryptography, where protocols are developed that use properties of the quantum internet. A way to prevent quantum attacks on current quantum unsafe cryptographic protocols (RSA, AES)~\cite{Shor1994, Grover1996} is to replace message authentication and sender non-repudiation by their quantum equivalents. These protocols are \textbf{quantum authentication}~\cite{barnum2002authentication} and \textbf{quantum digital signatures}~\cite{gottesman2001quantum,puthoor2016measurement, Chen2017}. Additionally, in geographies where data retention is regulated or where particular data of customers is confidential (e.g. General Data Protection Regulation (GDPR) in the European Union), \textbf{quantum encryption with certified data deletion} at scale could simplify compliance and prevent mistakes~\cite{broadbent2020quantum}. 

Furthermore, as discussed in the Section~\ref{section:intro_qi}, at early stages of the quantum internet, classical \textbf{encryption with QKD} is an accessible technology that will continue to be relevant going forward~\cite{Ekert1991, Bennett2014}. Lastly, consideration for distributed algorithms is dealing with faulty participants in networked decisions and how to get to an agreement among them. This is known as the Byzantine agreement problem. With V2X interconnectivity, faulty or malicious data exchange can lead to safety issues~\cite{Malik2020, pekaric2021taxonomyattack}, requiring to use Byzantine fault tolerant protocols~\cite{chen2021byzantine, Cheng2021byzantine}. We therefore include that quantum communication can offer a faster version known as \textbf{fast Byzantine agreement}~\cite{fastQbyzantineagreement}.

\vspace{\baselineskip}
\begingroup
        \centering
        \paragraph{Quantum sensors}
        \mbox{}\\
\vspace{\baselineskip}
\endgroup
\label{sec:quantum_sensors}
When a quantum internet is rolled out, we see the possibility of being able to use a distributed network of quantum sensors. This can be in the form of \textbf{quantum internet-of-things (QIoT)}. Another known application is to enhance the precision of distributed \textbf{clock synchronization} using quantum resources~\cite{giovannetti2001quantum}. However, in this application we include the usage of classical technology that helps distribute sub-nanosecond level timing distribution~\cite{lipinski2011white}.
Separately, quantum sensors can precisely measure acceleration offering 50 to 100-fold improvements in sensor stability over their classical counterparts~\cite{Templier2022,salducci2024}. This is a crucial technology for autonomous driving that complements global navigational satellite systems (GNSS) in GNSS-denied environments (tunnels, cities) to enable inertia navigation. Thus, enhanced \textbf{position tracking} is a relevant application of a distributed network of quantum sensors~\cite{Battelier2016}.

\subsubsection{Quantum Internet Research Trends}
\label{section:qi_research_trends}
In this section we will define research trends, and perform a mapping of those trends to their technological maturity in the quantum internet domain. We will use this maturity mapping to determine whether there is a relevant interface between the quantum internet trends and design considerations in the automotive pre-PDP requirements and discuss the results in Section~\ref{section:qi_prePDP}.

In defining the research trends, we aim to cover the entire quantum internet domain in three categories:
 \begin{enumerate}[noitemsep, label=\alph*.]
     \item \textit{Theory}
     \item \textit{Hardware}
     \item \textit{Software}
 \end{enumerate}
 The boundary between all three categories is fluid. In the category \textit{theory}, we recognize protocol development, analytical/numerical modeling of algorithms and projected hardware performance. This is distinguished from \textit{software}, which we categorize as development of software and architecture design that can directly be applied to code bases, libraries or specifically to (drivers of) software-controlled hardware. Lastly, with \textit{hardware} we cover the qubit platforms and classical control systems that control and interface with the quantum hardware. 
 
The trends that we recognize in these three categories are shown in Figure~\ref{fig:qi_megatrends} based on expert knowledge and global quantum internet research. While this list is carefully selected, it is non-exhaustive and exact trend recognition can vary by expertise.  We recommend revisiting this list periodically to ensure its relevancy.

We have mapped projected relevancy of each quantum internet megatrend to the functionality scale as defined in~\cite{Wehner2018}, shown in Figure~\ref{fig:qi_megatrends} on the horizontal axis. We start at the level of entanglement generation, a necessary basic task in quantum networking. If we extend such a network with few fault tolerant qubits we enhance the network's capabilities with small quantum computations and quantum memory. The last stage is to have full-fledged quantum computing mediated by a quantum internet, allowing arbitrary distributed quantum computing over multiple hardware systems. This range defines the area where quantum internet applications beyond (trusted node) QKD lie. Although all trends described are current topics of research and considerations, every trend has a different relevancy along the functionality levels of the quantum internet. To take one example, even though wafer-style fabrication is an important consideration at the early stages of building quantum internet technologies, these considerations only become significantly relevant when the qubit-platform has matured to be at least few-qubit fault tolerant.

We perform this mapping specifically to functionality level to be platform agnostic. Some quantum hardware platforms are further developed in their functionality level than others, however the underlying trends and the functionality relevance remain the same. In the following sections we discuss the trends per quantum internet research domain.

\vspace{\baselineskip}
\begingroup
        \centering
        \paragraph{Theory}
        \mbox{}\\
\vspace{\baselineskip}
\endgroup

Towards the \textbf{scaling} of larger networks, more protocol and architecture development is required both for orchestration of entanglement distribution and scheduling and routing of entanglement generation requests~\cite{gauthier2023,gauthier2024}. In contrast to the classical internet, most of these concepts in the quantum internet space are unexplored. Furthermore, quantum systems give rise to the challenge of optimal node placement in networks~\cite{Rabbie2022}. Theoretical models can provide tools to model hardware requirements, aiding their development and provide \textbf{optimizations}. 
The same holds for building frameworks to \textbf{benchmark quantum network performance}~\cite{vardoyan2022quantum, inesta2023performance}. 
Separately we consider \textbf{cryptography} as active area of research. Continued research is relevant to discover various ways of verified blind quantum computing. The application of cryptographic protocols and developing novel protocols can help enable quantum internet use cases, as has been shown for verified blind quantum computing~\cite{Fitzsimons2017, Drmota2024_VBQC_trapped_ions}.

\vspace{\baselineskip}
\begingroup
        \centering
        \paragraph{Hardware}
        \mbox{}\\
\vspace{\baselineskip}
\endgroup

The quantum internet domain is in search for improved quantum devices, carrying the name of \textbf{next generation devices} - those that go beyond the current state of the art in qubit performance~\cite{Bradac2019, Son2020, Simmons2024}. Since we are focusing on quantum communication, we recognize the required enhancement in photon emission and detection, i.e. finding new quantum emitters and improved single photon detection~\cite{esmaeil2017single}. For improved connectivity and looking forward to integrated devices, quantum frequency conversion hardware is being developed and pushed to efficiency limits both in conversion and in its applications of entanglement generation (time-bin, polarization encoding)~\cite{Bock2018, Dreau2018, Leent2020, Schaefer2023, Geus_2023}. 

Looking forward, with increased functionality of quantum devices comes the necessity of \textbf{miniaturization} of photon manipulation towards integrated photonics~\cite{Sipahigil2016, Stas2022, Pasini2023, Parker2024}, and moving to on-chip classical control electronics similar to quantum computing~\cite{Park2021, Ruffino2022}. This is combined with the drive to scale up quantum devices - more qubits per surface area as well as streamlined production processes with standardized outcomes and tolerance definition. Quantum device operation is potentially simplified if they could operate closer to room temperature, however this is a trend we only see relevant in the scaling of quantum computing systems.

Current state-of-the-art quantum network systems report uptimes of 78\% without human operator over the course of 17 days~\cite{Stolk2022}. Compared to classical internet industry standards of 99.9\%~\cite{GaurabBose2023}, continued work on \textbf{automation} of various layers in quantum networks is a necessity. This includes automated characterization of quantum chips, automatic routing reconfiguration~\cite{chakraborty2019distributed} and adaptive recalibration of connection parameters (entanglement fidelity, rate, etc.) in mature networks.

Lastly, the quality and availability of \textbf{interconnectivity} of different quantum devices is trending. Both in developing terrestrial and orbital quantum internet infrastructure~\cite{Simon2017,ForgesdeParny2023}, but also more directly in the challenges of distributing timing and synchronization of quantum devices on a network. Interconnectivity bandwidth can be improved from singular photonic channels to multiplexing on photonic channels to achieve the quantum analogy to wavelength-division multiplexing~\cite{Reimer2016}. Additionally, the concept of a quantum network controller is only in its infancy due to state-of-the-art quantum networks having at most several nodes, simplifying the control architecture. However, for many-node networks such a controller is paramount in delivering interconnectivity on demand. Finally, the physical transduction of quantum information encoded in different formats (photons, phonons, microwaves) is a building block in connecting different quantum hardware platforms~\cite{Weaver2024}. 

\vspace{\baselineskip}
\begingroup
        \centering
        \paragraph{Software}
        \mbox{}\\
\vspace{\baselineskip}
\endgroup

\textbf{Standardization} of control and software frameworks is essential for ensuring hardware interoperability, reducing costs, and simplifying maintenance. This includes agreement on a quantum network stack, of which initial attempts have been made already~\cite{dahlberg2019link, Dahlberg2022}. We recognize that the development of a quantum network stack is broad in itself, however it suffices for this analysis to group these developments under one trend. In standardization, there will be a requirement to continue updating standards as ISO or German Institute for Standardization (DIN), which is only relevant at higher maturity of quantum systems.

Especially standards aimed at \textbf{cybersecurity} and information management systems will keep up with the advancement of quantum technology. This makes certifying and auditing of quantum soft- and hardware a relevant trend in the long term, and the development of standards in the near term according to security-by-design principles. 

Finally, \textbf{adoption} of quantum communication tools is driven by the accessibility of simulation packages, both for software integration and application analysis~\cite{Coopmans2021, Wu2021, QNE_ADK_2024}.

\subsection{Megatrends Automotive}
\label{section:megatrends_automotive}
Megatrends across the entire automotive sector are defined by umbrella automotive organizations. Without loss of generality we focus on the German umbrella association (VDA) and their defined megatrends up to a timeline of 2030~\cite{VDA_megatrends_2021, VDA_roadmap_2021}. We extend that timeline towards $\geq 10$ years from publishing of this paper, given its continuous relevancy. 
 
First, we will discuss these future trends and then introduce the previously mentioned pre-PDP requirements. Then, in Section~\ref{section:automotive_future_trends_qi_applications}, we present the outcomes of the interfacing analysis between the automotive megatrends and quantum internet future applications.

\subsubsection{Automotive Future Trends}
\label{section:megatrends_autonomous_future_trend}
The VDA defines the following megatrend categories:
\begin{enumerate}[noitemsep, label=\alph*.]
    \item \textit{Autonomous driving and connectivity}
    \item \textit{Infrastructure}
    \item \textit{Mobility and logistics}
    \item \textit{Production}
    \item Materials
    \item Drivetrain and vehicle
    \item Energy carrier and storage
\end{enumerate}
We highlight in italic the megatrends we will discuss in this paper. The selected megatrends are those that actively use the classical internet, making them relevant to the quantum internet future applications and research tracks. The other megatrends related to material science and mechanical movement do not have a predicted significant relevance to quantum interconnectivity. From a quantum technology point of view, they would be more relevant to quantum computing (quantum chemistry, optimization)~\cite{AspuruGuzik2005,Hastings2015,bmwquantum2024}.

In the following paragraphs we give details about the above selected megatrend categories and their underlying trends. Additionally, for each trend we developed several use cases without assessment of their business viability. This allows the possibility for future cross-interaction of use cases between the automotive and quantum internet domains, even though their business viability might be uncertain today. The complete list of megatrend categories, trends and  use cases is shown along the vertical axis of Figure~\ref{fig:automotive_megatrend_qi_application}.

\vspace{\baselineskip}
\begingroup
        \centering
        \paragraph{Autonomous driving and connectivity}
        \mbox{}\\
\vspace{\baselineskip}
\endgroup

The main megatrend for the future of automotive is \textbf{autonomous driving}. As defined by the Society of Autonomous Engineers (SAE), there are 6 levels of automation in driving. They range from L-0 being solely the human as driver to L-5 where the vehicle is able to drive autonomously in any environment without failure~\cite{sae_autonomous_driving}. Many different components and yet to be developed technology need to be integrated in vehicles and surrounding infrastructure for L-5 to be achievable, of which we recognize several trends.

First, the act of \textbf{cooperative driving} together with \textbf{secure V2X communication} is an emerging field within autonomous driving, giving rise to the need for vehicle ad-hoc networks (VANETs) - vehicles only need interconnectivity with each other if actual interaction is expected , e.g. at an intersection~\cite{vanet_Farsimadan2021}. These ad hoc networks require self-organization, real-time communication and secure (anonymous) transmission, to mention a few features.. For more organized and higher abstraction level connectivity, \textbf{network management} is necessary for e.g. latency control. Vehicles will also need to be able to perform \textbf{situational prediction} of previously unknown situations and ensure the data they receive is correct. They are mission driven: a vehicle must successfully assess situations and be able to autonomously make decisions and execute actions on the road, while remaining functionally safe for the driver and other traffic participants. Data collection of the environment is performed through \textbf{sensors and fusion of sensor data} which needs to be processed in real-time~\cite{Campbell2018}.  Lastly, selective data collected necessary in order to perform autonomous movements will have to be private and treated securely, making \textbf{data privacy \& ownership} a trend in this category.

\vspace{\baselineskip}
\begingroup
        \centering
        \paragraph{Infrastructure}
        \mbox{}\\
\vspace{\baselineskip}
\endgroup
\label{sec:infrastructure}
Together with the electrification transition of vehicles and the requirement to be more interconnected, existing infrastructure needs changes to support these transitions. A trend in infrastructure development is combining the car as battery with the \textbf{energy grid}, which involves organization of load-balancing systems for charging, vehicle-to-load (V2L) and vehicle-to-grid (V2G) systems that all have to comply with cybersecurity standards. Additionally, infrastructure needs to support cars communicating with each other and the (quantum) internet via digital or quantum channels which we denote by \textbf{physical communication}. 

\vspace{\baselineskip}
\begingroup
        \centering
        \paragraph{Mobility}
        \mbox{}\\
\vspace{\baselineskip}
\endgroup

Collaboration of vehicles in traffic is the separate trend of mobility. Developments in this trend mostly focus on \textbf{swarming, fleeting and platooning} of vehicles within intermodal traffic management.

\vspace{\baselineskip}
\begingroup
        \centering
        \paragraph{Production}
        \mbox{}\\
\vspace{\baselineskip}
\endgroup

Separately from how a vehicle operates once produced, we see relevant trends in the production facilities and the management of factory processes. Manufacturers are strongly interested in constantly reducing their resources and costs by increasing the efficiency of the production. Value chains and assembly lines are aimed to be optimized by making use of the Internet of Things through the digital connection of machines, buildings, and plant locations, but also through the subsequent use of artificial intelligence and machine learning. Extensive connectivity in factory plants enable using swarm effects for production facilities. Machines could organize themselves in a swarm to share information and learn from each other~\cite{plus10_machine_learning_production}. With more automated production, coordination of intelligent machines and timing needs to be synchronized. Furthermore, sources of data that are being shared within the factory need to be verified, i.e. data ownership requires guarantees. Furthermore, we see a continued need for secure data tracking during the entire production chain and secure interfacing with the factory network as trend.

\subsubsection{Pre-PDP Requirements}
\label{sec:megatrends_automotive_prePDP}
In this section we discuss the pre-PDP requirements that are suited for assessing interfacing relevancy with the current research tracks of the quantum internet. 

\vspace{\baselineskip}
\begingroup
        \centering
        \paragraph{Vehicle \& infrastructure}

        \mbox{}\\
\vspace{\baselineskip}
\endgroup

We define this category of requirements as system requirements for products that are aimed to be included in a vehicle and the (out-of-vehicle) infrastructure necessary to support it. In order to assess interfaces with other products, a \textbf{characterized performance} of each integrated product needs to be known and verified. These parameters are well defined in many kinds of product requirement documents (PRDs). Furthermore, the smaller the product can be, the better. A car has limited volume, and this makes \textbf{miniaturization} of products essential. Then, to increase the possibility of interfacing, replacing or  upgrade parts, the \textbf{modularity} of the product must be taken into account. Lastly, the ease with which products are set up is its \textbf{deployability}.

\vspace{3\baselineskip}
\begingroup
        \centering
        \paragraph{Regulations}
        \mbox{}\\
\vspace{\baselineskip}
\endgroup

With the automotive being a highly regulated sector with different rules per country, all products need to pass externally imposed regulations and laws. Additionally, they need to adhere to internal regulations that are generally manufacturer specific. We focus on the regulations that matter for this analysis, being \textbf{cybersecurity} in the form of UNECE R155\footnote{UN Regulation No. 155: Cyber security and cyber security management system~\cite{UNR155}}/R156\footnote{UN Regulation No. 156: Software update and software update management system~\cite{UNR156}}, 
ISO 27001\footnote{\textit{Information security, cybersecurity and privacy protection}~\cite{iso27001}}, 
ISO/SAE 21434\footnote{\textit{Road vehicles - Cybersecurity engineering}~\cite{iso21434}} and the recently introduced EU Digital Operational Resilience Act, to name a few relevant regulations out of many that apply to this sector. In addition, we recognize the importance of compliance with \textbf{energy consumption} regulations both for fossil-fueled and electric vehicles. One of the key requirements that distinguishes the automotive from most other consumer product fields is the need for technology to be \textbf{resilient against extreme environments}. Semiconductor technology in a vehicle can only be used if it is operable in almost any environment on earth and with road conditions that cause vibrations and even corrosion (ISO 16750\footnote{\textit{Road vehicles - Environmental conditions and testing for electrical and electronic equipment}~\cite{iso16750}}). Some components require extreme temperature range resistance from \SI{-55}{\celsius} to \SI{+125}{\celsius} due to a combination of low and high temperatures generated by both the outside environment and heat generation of the vehicle itself~\cite{iso16750,Seeck_2021}. Furthermore, products require a well assessed safety risk during crash conditions. These types of requirements are rarely considered in the development of quantum technologies as a whole, making it a highly relevant topic for interfacing analysis.

\vspace{\baselineskip}
\begingroup
        \centering
        \paragraph{Quality}
        \mbox{}\\
\vspace{\baselineskip}
\endgroup

Product quality must be consistent and thus \textbf{reproducible} within specified \textbf{tolerances} and should also retain its quality during production and after. Those requirements are product and original equipment manufacturer (OEM) specific, and will also be described in the product requirement documents. 

\vspace{\baselineskip}
\begingroup
        \centering
        \paragraph{Long-term availability}
        \mbox{}\\
\vspace{\baselineskip}
\endgroup

In the product requirement documents the long-term availability of the product must be stated, as vehicle product lines can run for years and thus \textbf{manufacturing continuity} has to be preserved. Additionally, the raw materials or parts used require \textbf{supply-chain availability}, which for quantum devices becomes relevant when rare-earth or other highly specific materials are used~\cite{Mans2023, Mans_comm_2023}.

\vspace{\baselineskip}
\begingroup
        \centering
        \paragraph{Change management}
        \mbox{}\\
\vspace{\baselineskip}
\endgroup

During the (pre-)PDP the possibility of making changes to the products will have to be possible to mediate interfacing. After production, the \textbf{product service life} must correspond to the service life of a vehicle. Additionally, firm- and software updates in the form of  \textbf{OTA updates} to products will have to be possible for the whole life cycle of $\geq10$ years.

\section{Interfacing Analysis}
\label{section:interfacing_analysis}
We now turn to the interfacing analysis, divided into three parts. First, we interface the pre-PDP requirements to the quantum internet trends in Figure~\ref{fig:qi_trend_prePDP} and discuss the results in Section~\ref{section:qi_prePDP}. Then we compare these outcomes with the maturity mapping of Figure~\ref{fig:qi_megatrends} to obtain a synergy evaluation of the automotive and quantum internet trends in Figure~\ref{fig:qi_automotive_synergy} and Section~\ref{section:synergy_evaluation}. Lastly, we interface the future trends of the quantum internet to currently foreseen use cases of the automotive future trends in Figure~\ref{fig:automotive_megatrend_qi_application} and discuss the use cases in Section~\ref{section:automotive_future_trends_qi_applications}. 

The structure of the following interfacing figures is a framework to elucidate mutual influences at different trend timelines: pre-PDP versus quantum internet megatrends in Figure~\ref{fig:qi_trend_prePDP} and automotive use cases versus quantum internet applications in Figure~\ref{fig:automotive_megatrend_qi_application}. Because of the mutual influence of the domains on each other, these figures are to be read bidirectionally.

\subsection{Quantum Internet Research Trends \& Pre-PDP Requirements}
\label{section:qi_prePDP}
\begin{figure*}
    \centering
    \includegraphics[width = 0.8\textwidth]{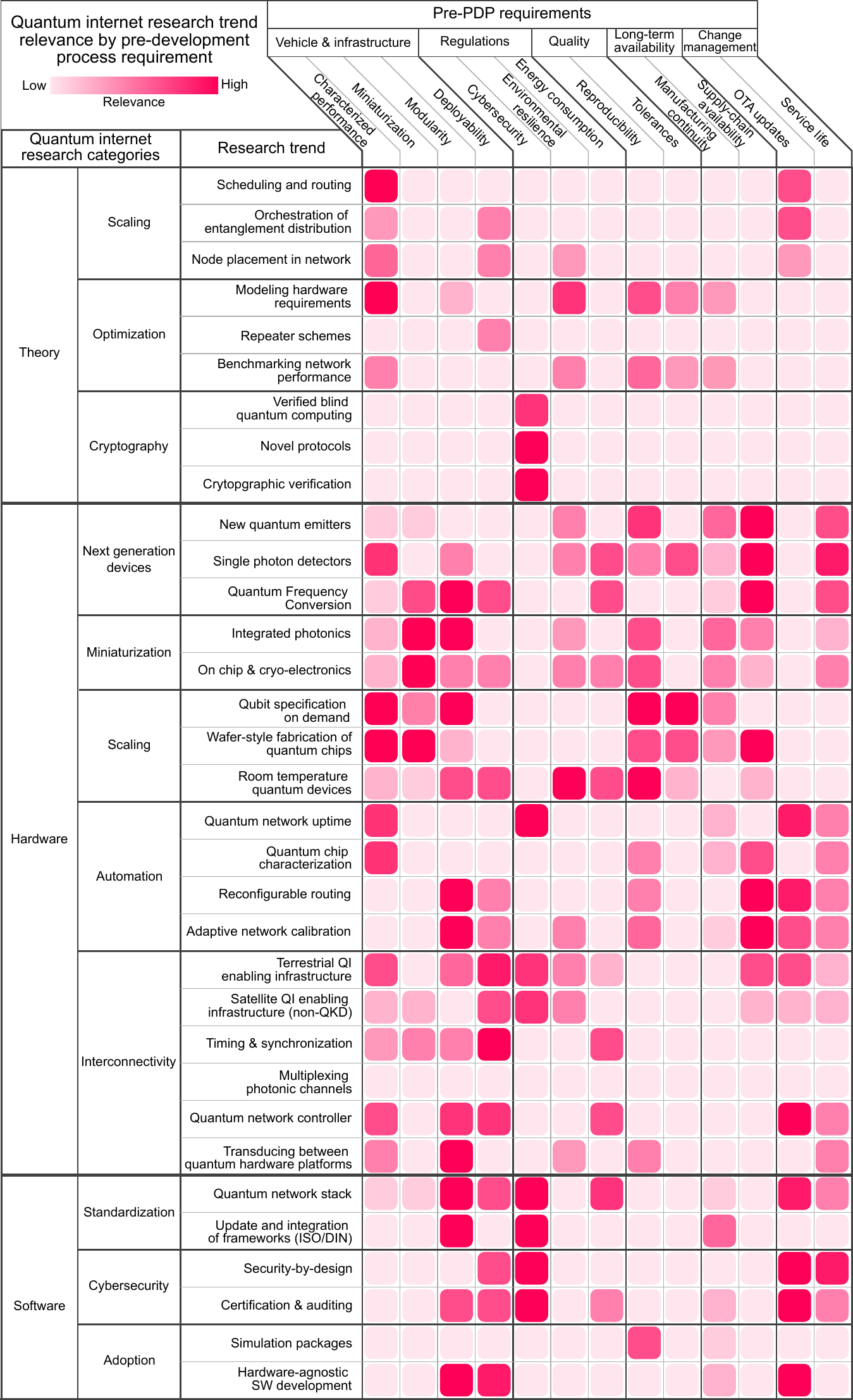}
    \caption{\textbf{Quantum internet and pre-PDP requirements analysis.} We evaluate the relevancy of quantum internet research trends and the categorized pre-PDP requirements. Relevancy of the trend intersection can be read bidirectionally. Note that there are more pre-PDP requirements than evaluated here, only the most relevant requirements for this analysis have been selected.}
    \label{fig:qi_trend_prePDP}
\end{figure*}

In this section we will describe the outcomes of the interfacing analysis of the quantum internet research trends and the pre-PDP requirements, shown in Figure~\ref{fig:qi_trend_prePDP}. We list the research trends of the quantum internet that were previously discussed in Section~\ref{sec:megatrends_QI} and shown in Figure~\ref{fig:qi_megatrends} on the vertical axis. On the horizontal axis we list the pre-PDP requirements from Section~\ref{sec:megatrends_automotive_prePDP} and together evaluate the bidirectional relevancy of these trends to each other. We define relevancy to be low when we evaluate the trends to have no interface with each other, they exist completely parallel to each other. The relevancy is high when the trends have a high degree of interfacing: they can mutually benefit from expert knowledge or requirements from the other domain. 
In the following sections we will discuss the interfacing of the pre-PDP requirements per quantum internet megatrend category. 

\subsubsection{Theory}
Trends in quantum internet theory developments are largely disconnected from production requirements. We can state that the megatrends of optimization and cryptography are a layer before being considered as pre-PDP requirement. Quantum cryptography is solely related to cybersecurity aspects of pre-PDP requirements, as they will have to comply to regulatory standards. This interface is one-directional, only the pre-PDP requirements can provide input for quantum cryptography developments, but the outcome of this research likely has little influence on the demands from the automotive side. Lastly, the availability of characterized performance is requested through the product requirement documents and thus modeling of hardware requirements and network performance will be essential.

\subsubsection{Hardware}
The highest relevancy in interfacing can be found in the hardware megatrends, which is a natural overlap given that the pre-PDP requirements are largely hardware focused. 

\vspace{\baselineskip}
\begingroup
        \centering
        \paragraph{Next-gen \& miniaturization}
        \label{section:prepep_vehicle_infra}
        \mbox{}\\
\vspace{\baselineskip}
\endgroup

Hardware megatrends directly relate to product integration into a vehicle. It is a necessity to have characterized performance of quantum internet hardware, as the automotive will demand clear performance requirements. This is valuable knowledge for hardware development strategies: the automotive can co-determine, with their use-cases, the minimum viable performance for the quantum connection, e.g. in required entanglement generation rate, fidelity, memory coherence and uptime. Additionally, the development of next generation quantum devices needs to take manufacturing continuity and supply-chain availability into account, which is highly relevant as several promising qubit systems are derived from rare-earth metals or diamond~\cite{Bradac2019}. 

Miniaturization is a trend present in both the pre-PDP and quantum internet research, thus interfaces well. The automotive can provide significant guidance on the miniaturization requirements. Besides `the smaller the better'always taking preference, together both domains can determine the threshold of trading off functionality of new quantum internet hardware with respect to its integration complexity. The efforts towards miniaturization on the quantum internet hardware are already largely in line with the automotive demands: especially integrated photonics and on-chip electronics provide order of magnitudes reduction in packing volume of both quantum hardware and classical control technology compared to their current free-space or in-fiber counterparts. Furthermore, miniaturization could reduce overall energy consumption, especially relevant for electric vehicles.

\vspace{\baselineskip}
\begingroup
        \centering
        \paragraph{Scaling \& automation}
        \mbox{}\\
\vspace{\baselineskip}
\endgroup

The scaling of quantum hardware systems will help satisfy pre-PDP requirements. Standardized qubit specifications and mass-producing quantum chips will ensure reproducibility and be within set tolerance requirements in quantum performance. This will also aid in determining compatibility with car control hardware. With improved specificity of device capabilities, (quantum) control hardware can be miniaturized further and integration can be better defined. The same arguments hold for more standardized fabrication methods. Together they can also guarantee manufacturing continuity and reproducibility of the product, especially when considering the production of 93 million vehicles yearly worldwide~\cite{oica_2023}.

In quantum internet automation we recognize interfaces in pre-PDP modularity and reconfigurability. Automated calibrations and rerouting can help to guarantee network uptime, even if parts of the quantum network are (temporarily) offline or unavailable. The hard- and software that organized this reconfiguration will have to be OTA update compatible and also should receive (security) software updates/support, hence the interfacing with the pre-PDP service life requirement.

\vspace{\baselineskip}
\begingroup
        \centering
        \paragraph{Interconnectivity}
        \mbox{}\\
\vspace{\baselineskip}
\endgroup

With the introduction of the internet and connecting vehicles to a telecommunications infrastructure, collaborative efforts were made to integrate vehicles computer systems with those infrastructures for navigation, in-vehicle-entertainment and OTA updates. We see a possibility in extending these efforts even further towards integrated quantum internet systems not only for the vehicles itself, but also for designing and setting requirements for the terrestrial and orbital quantum internet infrastructure and its future network controllers. In that context, the interconnectivity trends interface with most pre-PDP requirements pertaining to vehicle \& infrastructure and regulatory compliance.

\subsubsection{Software}

The standardization of both hardware and software related to quantum internet technology results in this field producing its own standards corresponding to their newly developed products~\cite{qirg}. The output of this megatrend thus becomes an input for the regulatory pre-PDP requirements, where required standards can now include quantum related standardization as well. 

This is opposite for cybersecurity, as this has become a cornerstone and topic weaved in almost every digital aspect of the automotive. Here the quantum internet domain has to adapt: clear guidelines and standards describe the security practices, auditing and complying to already existing standards. For any software to be deployable in a car, security-by-design and certification will have to be implemented. The progression from academic and research software to a software product can thus learn from the requirements set by the automotive, which is in part based on governmental (e.g. GDPR and Network and Information Security Directive (NIS2) in the European Union) or industry standards defined by e.g. ISO or DIN. 

Although adoption of quantum internet related software is an important megatrend, we see little relevance to the pre-PDP requirements. We do recognize that long life maintainability of software products can be achieved with more certainty  if software is continuously developed and maintained by many users and developers. A broad adoption of (open-source) code-bases can aid in achieving that goal, while allowing transparency around security protocols being implemented.

\subsection{Synergy evaluation}
\label{section:synergy_evaluation}
\begin{figure}[t]
\begin{minipage}{1\linewidth}
    \raggedright
    \includegraphics[width =0.811\linewidth]{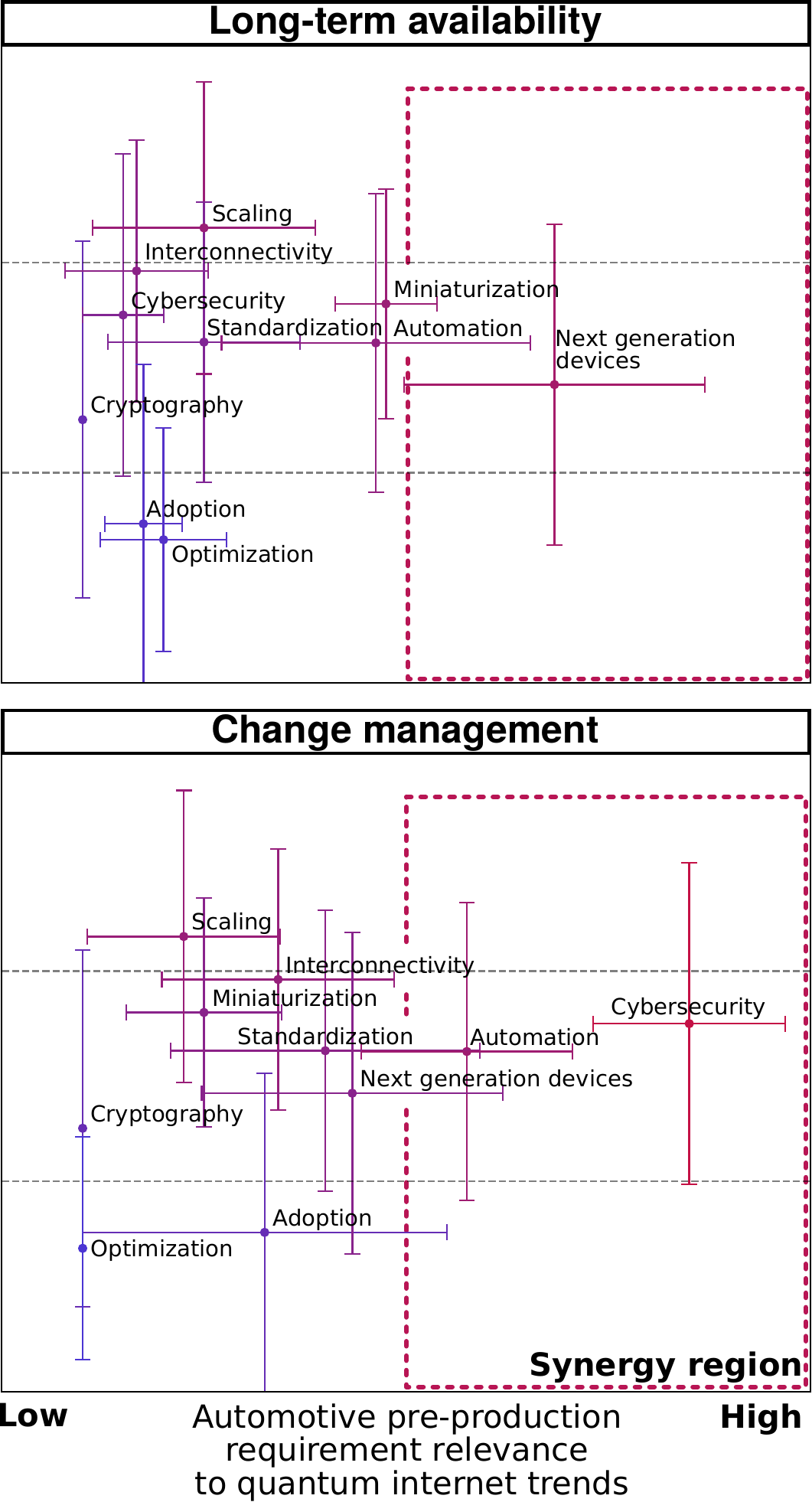} 
     \caption{\textbf{Synergy evaluation}:  We evaluate the quantum internet megatrends of Figure~\ref{fig:qi_megatrends} and the relevancy of the quantum internet megatrends with the automotive pre-PDP requirements of Figure~\ref{fig:qi_trend_prePDP} to obtain a synergy mapping. For the quantum internet megatrends we group the research topics per subcategory and show when it will become relevant on the quantum internet functionality scale of the vertical axis. We perform the same mapping for the pre-PDP interfacing analysis on the horizontal axis based on the data collected in Figure~\ref{fig:qi_trend_prePDP} and evaluate every pre-PDP category separately as denoted at the top of each graph. The dots show the averages of the relevancy per subcategory grouping of Figure~\ref{fig:qi_megatrends} (vertical axis) and Figure~\ref{fig:qi_trend_prePDP} (horizontal axis). The bars signify the average range (vertical axis) and one standard deviation of the relevancy (horizontal axis) per subcategory grouping. The dotted lines frame the synergy region, the region of interest where the potential for synergy between the automotive pre-PDP requirements and the quantum internet trends is the highest. This region covers all functionality levels of the quantum internet which allows this mapping to remaining hardware agnostic and cover near term and long term synergies within the $<10$ year evaluation timeline.}
     \label{fig:qi_automotive_synergy}
\end{minipage}
\end{figure}

To further assess the viability of collaboration on the short term ($<10$ years) between the automotive sector and the quantum internet, we perform a synergy evaluation as shown in Figure~\ref{fig:qi_automotive_synergy}. There we evaluate two parameters: 

\begin{enumerate}[noitemsep]
    \item Where in functionality level is each quantum internet megatrend relevant? (vertical axis)
    \item How relevant is that megatrend for the pre-PDP requirements? (horizontal axis)
\end{enumerate}
This combination allows us to map which quantum internet research trends have the ideal combination of being both near-term relevant in their development and highly relevant as pre-PDP requirements. It is those trends that allow for two-way exchange of expertise and requirements. We highlight the synergy region, the region in which we expect to find the largest potential for synergy between the automotive and quantum internet domains.
We can interpret the bottom of the synergy region to be the near term use case opportunity and the top mostly long term, however all to be within the $<10$ year timeline given by the pre-PDP requirements. 

We are only able to perform the synergy evaluation to the pre-PDP requirements, as they are evaluated against the quantum internet megatrends, of which we both have detailed assessments. Such an assessment is not yet possible to be made for the automotive megatrends as the quantum internet applications are on the $\geq10$ year timeline. We discuss the topics that are part of the synergy region in the following sections.

\subsubsection{Synergy regions}

\begingroup
        \centering
        \paragraph{Vehicle \& infrastructure}
        \mbox{}\\
\vspace{\baselineskip}
\endgroup

 Miniaturization has significant potential for synergy with the pre-PDP vehicle and infrastructure requirements. As mentioned in Section~\ref{section:prepep_vehicle_infra}, we identify short-term interfaces in miniaturization:
 
\vspace{0.2em}
\begin{center}
\begin{tabular}{ l l }
 $\bullet$ Form factor &  $\bullet$ Modularity  \\ 
  $\bullet$ Integration complexity &  $\bullet$ Power consumption \\  
  $\bullet$ Environmental resilience & \\    
\end{tabular}
\end{center}

\vspace{\baselineskip}
\begingroup
        \centering
        \paragraph{Long-term availability}
        \mbox{}\\
\vspace{\baselineskip}
\endgroup

From a hardware perspective, it is vital that the next generation devices of the quantum internet support long-term availability of its used raw materials and production processes. This is especially relevant for the short-term, as many new quantum devices are actively being developed to be used in quantum networking as communication qubit.

\vspace{\baselineskip}
\begingroup
        \centering
        \paragraph{Regulations}
        \mbox{}\\
\vspace{\baselineskip}
\endgroup

A natural synergy is found in the development of standardization in general and cybersecurity, specifically. The automotive can provide input on these topics in advance of any quantum internet related product entering the market. Clear stated requirements from regulations can provide a smooth transition of new quantum technology from experimental to a commercial vehicle-deployable one. To ensure this transition, it is beneficial  that quantum product certification will align with e.g. ISO 27001, 21434 or 16750 standards. Conversely, the automotive is highly dependent on the development of the quantum network stack and will need to adapt their systems based on the requirements that this stack demands of their hard- and software. For instance, to operate the quanutum network stack, the availability of particular antennas for communication functionalities could necessary. Additionally, specific requirements on the vehicle's real-time computing power and data transfer bandwidths are also to be expected.

\clearpage

\begin{figure}[t]
\begin{minipage}{1\linewidth}
    \centering
    \includegraphics[width =1\linewidth]{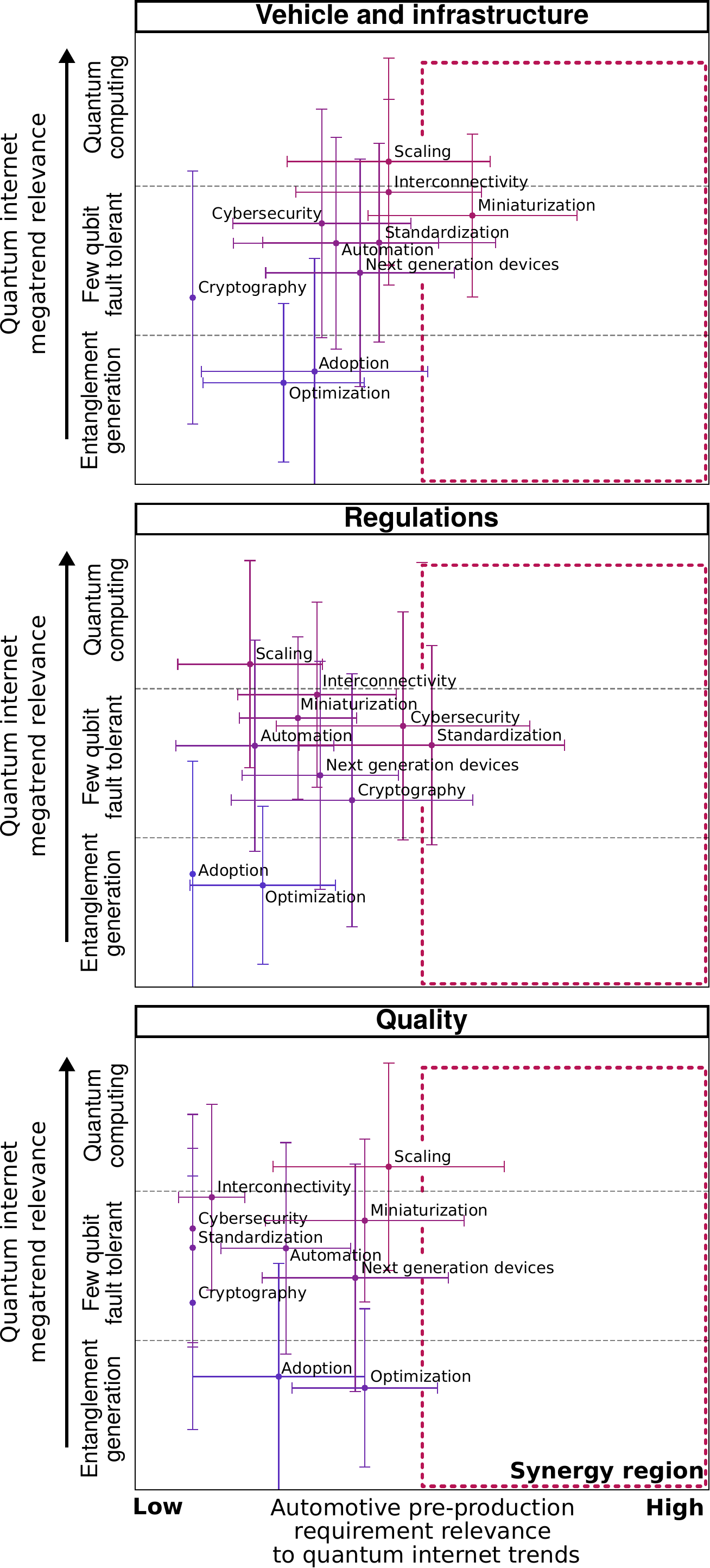}
\end{minipage}
\end{figure}

\clearpage
\vspace{\baselineskip}
\begingroup
        \centering
        \paragraph{Change management}
        \mbox{}\\
\vspace{\baselineskip}
\endgroup

We find strong synergy for the quantum internet domain to incorporate cybersecurity in their products, matching with the same requirement for vehicles.   as well as enabling software-powered automation of hardware systems. It is much easier to integrate security-by-design in fresh code bases for new products. Certification demands and operational automation are additionally better to be integrated at the start of a product development process than after. 
These product development considerations are a hard requirement for the eventual breakthrough of a quantum internet product in any mature industry.

\vspace{\baselineskip}
\begingroup
        \centering
        \paragraph{Quality}
        \mbox{}\\
\vspace{\baselineskip}
\endgroup

There is a small opportunity for quantum internet hardware scaling to be relevant in quality requirements at pre-PDP. Especially in reproducibility and tolerances the automotive can provide their input on qubit parameters. Other trends of this category are not sufficiently in the synergy region.

\subsection{Automotive Future Trends \& Quantum Internet Applications}
\label{section:automotive_future_trends_qi_applications}

\setcounter{paragraph}{0}

\begin{figure*}
    \centering
    \includegraphics[width = 0.8\textwidth]{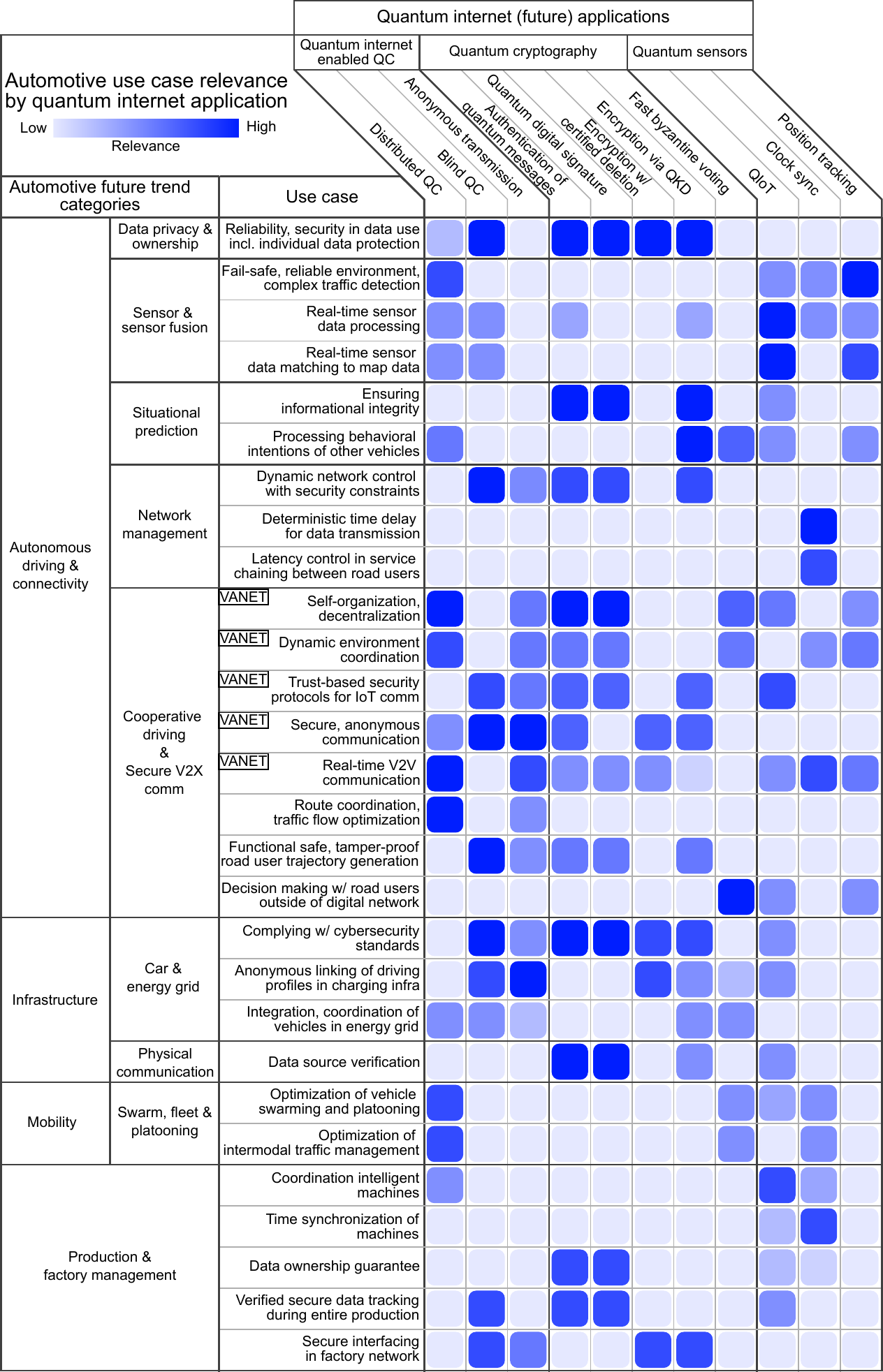}
    \caption{\textbf{Automotive megatrends, use cases and the quantum internet future applications.} We list the automotive future trends (vertical axis) as discussed in Section~\ref{section:megatrends_autonomous_future_trend}, defined by the German umbrella association for automotive~\cite{VDA_megatrends_2021}. We developed several use cases per trend and evaluate the relevancy of those use cases to quantum internet future applications (horizontal axis). The relevancy can be read bidirectionally for each trend. }
    \label{fig:automotive_megatrend_qi_application}
\end{figure*}

The final interfacing analysis we make is of the future trends of the automotive, defined in Section~\ref{section:megatrends_autonomous_future_trend}, with the quantum internet future trends of Section~\ref{section:qi_applications}. We developed a list of potential automotive use cases out of the future trends through a structured brainstorming approach of challenge-solution fits~\cite{meinel2011design, Wilms2022}. 
The challenges are developed from the automotive megatrends and solutions are proposed by the future applications of the quantum internet. 

The automotive use cases are shown on the vertical axis of Figure~\ref{fig:automotive_megatrend_qi_application}, and we assessed the potential of applying quantum internet applications (horizontal axis) to these use cases.

\vspace{\baselineskip}
\begingroup
        \centering
        \paragraph{Autonomous driving and connectivity}
        \label{section:autonomous_future_trend}
        \mbox{}\\
\vspace{\baselineskip}
\endgroup

The autonomous driving and connectivity megatrend has relationships across the entire range of quantum internet applications. Use-cases pertaining to optimization of routes and traffic are both related to distributed quantum computing as well as enhanced sensing: having more quantum computing power available in a distributed fashion can allow for improved calculations on efficient traffic flows. Especially the creation of VANETs can benefit from blind quantum computing and anonymous transmission in communication and complex decision processing (see Sec.~\ref{sec:interconnected_quantum_computing}). The quantum communication protocols we already know to provide enhanced privacy, security or anonymity are potentially relevant for all automotive use cases related to communication or security. 

Furthermore, local autonomous decisions can be aided with enhanced sensory input from quantum sensors. As previously discussed, quantum accelerometers can aid GNSS-denied areas by enabling inertia-navigation (see Sec.~\ref{sec:quantum_sensors}), offering a substantial enhancement of accelerometer stability 50-100 fold over their classical equivalent. Other quantum sensors can provide enhanced light detection and ranging (LiDAR), offering a factor 4 theoretical improvement in signal-to-noise ratio compared to optimal classical radar and a theoretical factor $\sqrt{2}$ improvement in precision for target distance measurement~\cite{Slepyan2022}. This topic can be combined for sensors to enable high precision multi-parameter optimization.

Additionally we recognize the potential of fast Byzantine voting (see Sec.~\ref{sec:quantum_cryptography}) to aid local decision making between ensembles of vehicles interacting in traffic situations. This can allow for fewer-round decision making as well as preventing bad actors from disturbing traffic flow or reducing the safety of traffic participants~\cite{Malik2020, Liu2019}. Important to note is that, although the concepts of Byzantine voting are well developed within classical and quantum information theory, applications within the context of automotive as described specifically using quantum connectivity require the collaboration and input from both domains to find solutions that can be applied to real-life challenges. Such a collaboration benefits specifically from finding solutions that provide an improvement that can proven theoretically (from quantum network theory perspective) and has potential to be implemented practically (from the automotive perspective).

\vspace{\baselineskip}
\begingroup
        \centering
        \paragraph{Infrastructure}
        \mbox{}\\
\vspace{\baselineskip}
\endgroup

There are interfaces between the megatrend of infrastructure around electric charging/V2L/V2G (see Sec.~\ref{sec:infrastructure}) and cybersecurity with blind quantum computing, anonymous transmission, and all quantum cryptography protocols pertaining to authentication and verification (see Sec.~\ref{sec:quantum_cryptography}). We recognize that the quantum internet can provide tools to ensure anonymity. Quantum authentication and signatures can help provide data source verification resistant to quantum attacks. Fast Byzantine voting can aid in finding consents in a V2G connected network, as it is a network with many actors with fast and live changes on power demands, while ensuring privacy among cars and without requiring privileged members or leader election. This could be used for the coordination of vehicles at charging stations, even if one car shares misinformation.

Not mentioned in Figure~\ref{fig:automotive_megatrend_qi_application} are the reverse use cases where automotive technology can provide use cases for the quantum internet. Their implementation is too uncertain to consider in the relevancy analysis. For example, the concept of quantum edge computing becomes feasible with the appropriate infrastructure support from telecommunications and the automotive. Quantum processing tasks can be delegated in a blind or non-blind way to a more powerful quantum computer close by, while having a smaller quantum computer available locally in the vehicle that has the ability to teleport information to/from this edge quantum computer.

Complementary, we notice the possibility of vehicle-supporting infrastructure to provide quantum states either (1) statically through charging ports, where a car is loaded with quantum states to an integrated quantum memory or even (2) dynamically even while driving through free-space connections we shall denote as quantum antennas. Proposed research directions can include:
\begin{itemize}[noitemsep]
    \item Optimization of geographical positioning of quantum antennas within urban areas
    \item Protocol development and standardization of free-space communication architectures to distribute flying qubits
    \item Development of antenna hardware for multiple-party sending and receiving of single photons for entanglement generation
\end{itemize}

\noindent Beneficial to this cause is the development of the 6G standard, which conceptually attempts to incorporate the requirement that quantum networks could put on the classical internet, i.e. low-latency and incorporation of quantum-generated classical encryption keys. This trend will continue for communication standards beyond 6G, allowing for continued integration of quantum networks within communication standards.

\vspace{\baselineskip}
\begingroup
        \centering
        \paragraph{Mobility \& logistic concepts}
        \mbox{}\\
\vspace{\baselineskip}
\endgroup
    
The mobility \& logistic concepts megatrend focuses on new types of logistical concepts that can be applicable for management of vehicles and transport systems. As such, it is a topic that has considerably less overlap with quantum internet future trends. Only a few interfaces stand out: first, the organization and planning of swarming and platooning using distributed quantum computing resources. This can be performed with minimal (dynamical) input from the vehicles themselves. Second, vehicle-group alignment in decision making through fast Byzantine voting and clock distribution. This latter trend can benefit from focused collaborative research of quantum computer scientists and the automotive, as mentioned above in Section~\ref{section:autonomous_future_trend}, to identify whether or not there is an (theoretical) advantage in leveraging these quantum internet applications for swarms or platoons. 

Not mentioned in Figure~\ref{fig:automotive_megatrend_qi_application} are potentially new mobility concepts that include ridesharing and flexible car renting. These could potentially benefit from integration with cross-platform financing using a quantum money scheme\footnote{A quantum money scheme is a cryptographic protocol that uses the concepts of superposition and no-cloning to ensure its uniqueness and prevent counterfeiting.} to verify payments~\cite{wiesner1983,Gavinsky2012}, to securely verify identities using quantum digital signatures. Recently this has been shown to be applicable to e-commerce platforms~\cite{Cao2024}, which is conceptually similar to the transactions surrounding flexible car renting. However, while valuable ideas for future applications, these use cases are too uncertain in relevancy for the automotive and thus omitted from the interfacing analysis.

\vspace{\baselineskip}
\begingroup
        \centering
        \paragraph{Production}
        \mbox{}\\
\vspace{\baselineskip}
\endgroup

We specifically see a role for privacy preserving analytics in the digital and data trend of production, to guarantee the security and uniqueness of data processed and generated in production factories using quantum authentication and signatures.

\section{Conclusion}
We have presented a multi-layered use case analysis of the quantum internet's current research tracks and long-term applications applied to all facets of the automotive industry. We employed a method where cross-platform megatrend relevancy is qualitatively assessed. This method was successful in finding potential synergies between trends in the quantum internet and the automotive. Given the generality of the method it can further be applied to any arbitrary industry, allowing for a structured approach to cross-platform use case development of the quantum internet.

We compiled megatrends of the quantum internet based on current research tracks for a $<10$ year horizon and mapped their relevancy by level of quantum internet functionality. This comprehensive mapping is a first attempt at a hardware-agnostic trend analysis in the quantum internet domain. This will serve as foundation for further use case research for other industries than the automotive. With this mapping we have qualitatively evaluated which of these quantum internet megatrends have mutual relevance with pre-production engineering process requirements for the automotive sector. 

Together these two mappings combined to a synergy evaluation where strong synergy potential has been found in quantum internet hardware miniaturization and automation for vehicle \& infrastructure development, as well as for quantum internet software development on the quantum network stack to provide input on regulations for the automotive. Additionally, we find that supply-chain availability is a strong driver for allowing next generation quantum hardware to be included in the vehicle development cycle.

We evaluated megatrends on the $\geq10$ year horizon and produced a list of future use cases for the automotive. We assessed which quantum internet future applications are of potential interest to add value or functionality to the automotive use case. We have found that connected quantum sensors have the potential to serve in high precision multi-parameter optimization. Additionally, there is considerable potential for incorporating blind quantum computing, anonymous transmission and quantum authentication and signatures to augment the security and privacy aspects of V2G and autonomous driving, in particular in VANETs. 

\section{Recommendation for Action}
We have identified potential interfaces for quantum internet applications to be integrated into automotive products and processes. However, this analysis assumes optimal implementation conditions that may be difficult to achieve in practice. The deployment of these applications requires deep knowledge of quantum technologies as well as expertise on manufacturing, engineering production processes and more. Accumulating this knowledge requires a substantial investment both from the quantum internet and the automotive domain in research and development. Venturing into this shared domain poses significant strategic risks due to the many challenges and unclear feasibility of quantum technologies. Pursuing such an initiative is therefore only recommended for companies with a first mover strategy, or those willing to take high-risk, long-term endeavors. 

A possible route for the automotive industry to develop in this area is to build up know-how in the field of the quantum internet through academic collaborations. By leveraging those collaborations, the automotive industry can gain exposure to the quantum internet domain at an early stage without substantial investments. They can take a passive approach by observing the activities in the quantum internet domain, or choose a more active approach by participating in co-creating novel applications and communicate pre-development requirements. The earlier such a collaboration is started, the higher the probability that assimilation of quantum internet applications is eased in the automotive ecosystem. The close collaboration between the automotive industry and academic groups therefore occupies a crucial role in driving a shared development. In the academic context, the automotive pre-PDP requirements and knowledge about mass production could benefit the quantum internet field at this early stage. Venturing, cooperatively funded projects and commercial testbeds of relatively mature applications (miniaturization, environmental resilience, sensor integration) could further strengthen the intersection of these domains.

Lastly, the extensive analysis of this paper helped identify several fundamental research topics that are unexplored that provide interesting possibilities for an academic or start-up venture collaboration between the quantum internet and automotive: theoretical and algorithm development for quantum communication powered VANETs focused on fast Byzantine voting, adversarial-proof decision making and collaborative QIoT for multi-parameter optimization can be first steps towards predicting quantum advantage of quantum enabled interconnected vehicles.

\addtocontents{toc}{\protect\setcounter{tocdepth}{-1}}
\section*{Disclaimer}
\addtocontents{toc}{\protect\setcounter{tocdepth}{2}}
\noindent\textit{The results, opinions and conclusions expressed in this publication are not necessarily those of Porsche Digital GmbH or Audi AG.}

\section*{Acknowledgements}
\noindent We would like to thank Prof. Dr. ir. Ronald Hanson, Dr. Mahdi Manesh, Claudia Schwarz, Stefan Zerweck and Mattias Ulbrich for their support and the opportunity to perform this analysis. Dr. Clemens Wickboldt, Amirtharaj Amirthalingam and Dr. Sjoerd Loenen for the critical discussions and feedback on the manuscript. Laura Ohff, Dr. Marie-Christine Slater for useful discussions at the start of this project.\\

\noindent \textbf{Funding:} We acknowledge funding the Dutch Research Council (NWO) through the project “QuTech Part II Applied-oriented research” (project number 601.QT.001). We further acknowledge funding through the BMBF (HYBRID) and Porsche Digital GmbH.\\
\textbf{Competing interests:} The authors declare no competing interests.\\
\textbf{Data and materials availability:} The data supporting this manuscript is available at 4TU.ResearchData~\cite{manuscript_data}.\\
\clearpage

\bibliography{main.bib}

\end{document}